\documentclass[preprint2]{aastex}

\usepackage{amsmath}
\usepackage{graphicx}
\usepackage{epsfig,psfrag,epic,eepic}
\usepackage{textcomp}
\usepackage{times}
\usepackage{longtable}

\slugcomment{To be submitted to the Astrophysical Journal}

\shorttitle{Transiting Trojan Planets}
\shortauthors{Janson et al.}

\begin{document}

\title{A Systematic Search for Trojan Planets in the \textit{Kepler} data}

\author{Markus Janson\altaffilmark{1,2}
}

\altaffiltext{1}{Department of Astrophysical Sciences, Princeton University, NJ 08544, USA; \texttt{janson@astro.princeton.edu}}
\altaffiltext{2}{Hubble fellow}

\begin{abstract}\noindent
Trojans are circumstellar bodies that reside in characteristic 1:1 orbital resonances with planets. While all the trojans in our Solar System are small ($\lesssim$100~km), stable planet-size trojans may exist in extrasolar planetary systems, and the \textit{Kepler} telescope constitutes a formidable tool to search for them. Here we report on a systematic search for extrasolar trojan companions to 2244 known \textit{Kepler} Objects of Interest (KOIs), with epicyclic orbital characteristics similar to those of the Jovian trojan families. No convincing trojan candidates are found, despite a typical sensitivity down to Earth-size objects. This fact can however not be used to stringently exclude the existence of trojans in this size range, since stable trojans need not necessarily share the same orbital plane as the planet, and thus may not transit. Following this reasoning, we note that if Earth-sized trojans exist at all, they are almost certainly both present and in principle detectable in the full set of \textit{Kepler} data, although a very substantial computational effort would be required to detect them. On the same token, we also note that some of the existing KOIs could in principle be trojans themselves, with a primary planet orbiting outside of the transiting plane. A few examples are given for which this is a readily testable scenario.
\end{abstract}


\section{Introduction}
\label{s:intro}

The fourth and fifth Lagrangian points (L4 and L5) of a planet orbiting a star are stable, as long as the planet is less than $\sim$4\% of the system mass \citep[e.g.][]{murray1999}, which is practically always the case, since 4\% of 1~$M_{\rm Sun}$ is $\sim$40~$M_{\rm Jup}$. As a result, several of the planets in the Solar System have planetesimals in 1:1 orbital resonances around these points, collectively referred to as `trojans'. The largest known trojans have sizes of up to $\sim$100~km in radius \citep[e.g.][]{fernandez2003}, but there is nothing that physically prevents them from being as large as the Earth or even Jupiter in other planetary systems. While the stability criterion has traditionally been derived within the context of the restricted three-body problem \citep[e.g.][]{murray1999}, where the trojan is assumed to have zero mass, \citet{laughlin2002} have made the same calculation for the case where the trojan and planet have equal masses (in that context better referred to as a resonant planet pair), and found that the stability criterion remains essentially the same; it therefore appears that the distribution of mass between the planet and trojan is irrelevant, and that the relevant stability criterion is that sum of their masses is less than $\sim$4\% of the system mass. 

While some work has been performed in examining the possibility of existing trojans in known planetary systems and in trying to identify such objects \citep[e.g.][]{dvorak2004,gozdziewski2013}, none have so far been detected. Meanwhile, the \textit{Kepler} telescope has been collecting photometric data of $\sim$150,000 KIC (\textit{Kepler} Input Catalog) stars since 2009 in a search for exoplanetary transits, and has so far found more than 2000 objects of interest \citep[e.g.][]{batalha2013}, most of which are probably real planets \citep[e.g.][]{morton2011,santerne2012,lissauer2012}. Since the \textit{Kepler} data set is present, and has nearly continuous coverage of its targets and an exquisite photometric precision, it provides an opportunity to detect trojan companions to the already detected planet candidates in the data. A search for such objects however needs to be specifically dedicated for the purpose. Trojans often librate around their equilibrium points to a substantial degree, which leads to large transit timing variations \citep[TTVs; see][]{agol2005,holman2005,ford2007}, and thus they are easily missed in most planet-search algorithms, which require near-constant orbital periods in order not to miss most of the total signal. Furthermore, many such searches are based on periodograms, in which the trojan does not produce any distinct signal, since it shares a mean orbital period with its planet. For these reasons, we have performed a systematic search for extrasolar trojans in the \textit{Kepler} data, which we describe in this paper.

It is important to note that while there are good prospects for detecting trojan signals in the \textit{Kepler} data, there is not much that can be stringently constrained from their absence in case none are detected. The primary reason for this is that with \textit{Kepler}, we can only detect trojans with orbits that are co-planar with the planet. By contrast, the Jovian trojans have a rather broad distribution of inclinations, with up to $\sim$30$^{\rm o}$ deviation from Jupiter's orbital plane \citep{jewitt2000}. Hence, we cannot put global upper limits on the frequency of (sufficiently large) trojans from non-detections in the lightcurves, since such trojans may exist but are inclined with respect to the planetary orbit, and thus escape detection. We can only put constraints specifically on co-inclined trojans. 

Nonetheless, again, if there are co-inclined trojans in the \textit{Kepler} sample, we have the opportunity to detect them, and we present a search to this end in Sect. \ref{s:knownkois}. Furthermore, given that we know that trojans can have non-zero mutual inclination to their planets, we can entertain the thought that some of the known KOIs are not primary planets themselves, but trojan companions to a larger planet that is inclined to the line of sight and thus does not transit itself. This is particularly interesting for some KOIs that undergo clear TTVs, but where no other planetary signal has yet been observed in the system. We will discuss a few such cases in Sect. \ref{s:kois_themselves}. Finally, we will discuss some future prospects to study trojans, including possibly habitable trojans, in Sect. \ref{s:future}.

\section{Trojan Companions to Known KOIs}
\label{s:knownkois}

In this section, we describe the search for secondary signals from trojans in KOI systems -- i.e., the systems where a primary set of signals has already been detected and interpreted as a possible planet signal. Throughout this section, we will consistently make the following assumptions: Firstly, we will consider only systems for which the trojan is co-inclined with the planet, since otherwise we may not see its transit signature even if it physically exists in the system. Secondly, since the KOIs are all represented by a periodic signal that dominates the vicinity of its periodicity space and any trojan signature that may exist in the data is necessarily weaker, we are by default considering systems for which the trojan is significantly smaller than the planet. Since mass tends to scale rapidly with radius \citep[e.g.][]{weiss2013}, it is therefore also a natural assumption that it is significantly less massive. In this framework, we do not need to consider the semantic confusion that might arise if the two masses are similar, such that the intrinsic hierarchy of the terms `planet' and `trojan' becomes ambiguous. Here we will refer to the least massive body in the three-body problem as the `trojan', the middle mass as the `planet' or `primary planet', and the most massive body as the `star'. Obviously, it should be kept in mind throughout that most of the so-called `planets' have not been individually confirmed and should only be regarded as candidates.

In addition, we only consider trojans that have a libration amplitude of $<$23$^{\rm o}$ with respect to the star in a co-moving frame, similar to FWHM of the trojan population of Jupiter \citep{jewitt2000}. While larger amplitudes are probably also stable, and even horseshoe orbits are stable for sufficiently small masses \citep{laughlin2002}, they involve an increasing complexity of the orbit and a decreasing validity of the approximations used, and are substantially more computationally demanding to study, hence we restrict ourselves to the family of orbits for which a large population of objects are empirically known to be long-term stable. For reference, we have however kept an eye out for such large-amplitude and horseshoe orbits in our visual inspections of the data, but found no convincing candidates. Finally, we note that since we have assumed co-planar orbits, it naturally follows that we also assume equal transit durations $t_{\rm dur}$. With the exception of ingress and egress, the transit duration is independent on planet size, so the two most important parameters for the duration (given that the stellar size is obviously equal for the planet and trojan) are the impact parameter and the orbital velocity. The radial variations in a trojan orbit are negligible, so if the inclinations are equal, then the impact parameters are equal as well. Meanwhile, the epicyclic velocity is only $\lesssim$1\% of the orbital velocity, hence its impact on the duration can also be neglected.

\subsection{Procedure}

\subsubsection{Data Pre-processing}
\label{s:preprocessing}

\begin{figure}[p]
\centering
\includegraphics[width=8cm]{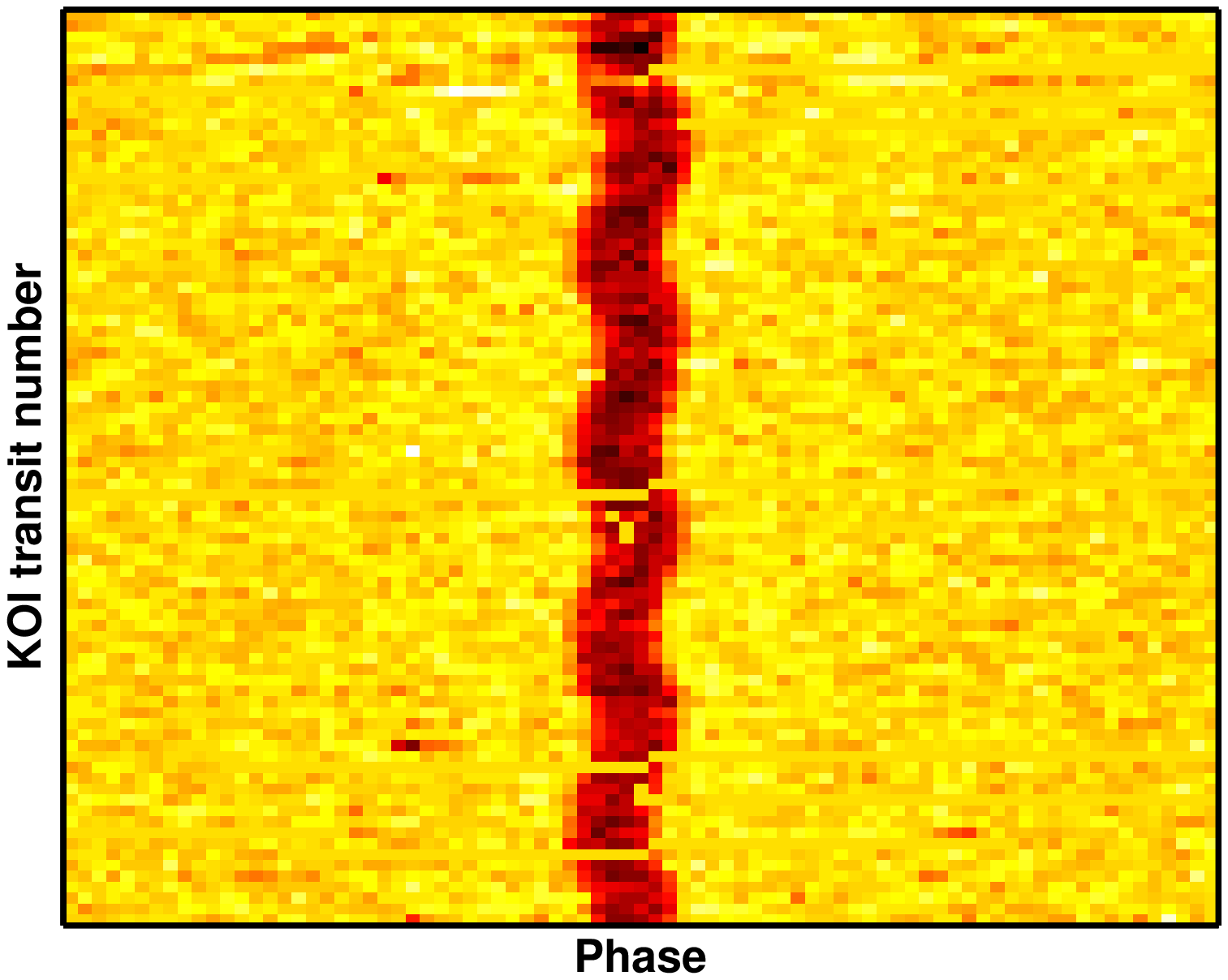}
\caption{General example of a river diagram. Shown here is a section of a diagram centered on a phase of 0.0, for the quasi-periodic transits of KOI-103.01. The approximately sinusoidal transit timing variations of KOI-103.01 creates a wiggled vertical trace along the diagram. The particular case of KOI-103.01 is discussed in more detail in Sect. \ref{s:kois_themselves}}
\label{f:river_example}
\end{figure}

A transiting trojan planet will manifest itself in a lightcurve as a set of quasi-periodic transits, with a mean period equal to that of the planet with which it shares a 1:1 resonance. If we define the mid-transit phase of the primary planet as $\phi_{\rm p} = 0$, then the typical trojan transit ephemeris will be at phases near $\phi_{\rm t} = \pi / 3$ or $\phi_{\rm l} = - \pi / 3$, corresponding to the 60$^{\rm o}$ by which the L5 and L4 Lagrangian points are trailing and leading the primary planet's orbit, respectively. The ephemeris will oscillate around these points due to the epicyclic motion of the trojan. As a result of this quasi-periodic behaviour, it is useful for the analysis and for illustrative purposes to structure the lightcurve in a `river diagram' framework, which has been previously used in the analysis of quasi-periodic orbits \citep{carter2012,carter2013}. A river diagram is a folded lightcurve in matrix form, in which each column represents a specific phase with respect to a fixed orbital period, and each row is one such period. A planet with a constant period will leave a straight vertical trace in a river diagram folded to its period, whereas a quasi-periodic planet will leave a wiggly vertical trace if the diagram is folded to its mean period. A general example of a river diagram is shown in Fig. \ref{f:river_example}. In our case, we create one river diagram per KOI by folding to the KOI period, and impose $\phi_{\rm p} = 0$ in practice by shifting the primary planet mid-transit phase to the first column of the matrix. Hence, in this framework, transiting trojan planets will be represented by wiggly traces near specific columns (corresponding to $\phi_{\rm t} = \pi / 3$ or $\phi_{\rm l} = - \pi / 3$) in the matrices.

Before converting the lightcurves into river diagrams, we subject the data to a subtraction of a median moving box filter with a box width equal to four times the transit duration. This eliminates most low-frequency variations, which benefits visual inspection, and does not significantly affect the strength of the transit feature (the specific morphology of the transit is affected, as it becomes slightly elevated with respect to the mean continuum flux, but since the continuum directly before and after the transit is elevated along with it, the strength and visibility of the feature itself is conserved). The lightcurves used were those produced by the standard \textit{Kepler} pipeline, for quarters Q0--Q12. The KOIs and their values were taken from the official \textit{Kepler} listing on March of 2013, and include those (and only those) KOIs that were either classified as planets or planet candidates at that time. This gives a total of 2244 KOIs, where the count includes individual KOIs in candidate multi-planet systems.

\subsubsection{Orbit Calculation}
\label{s:orbit}

As mentioned in Sect. \ref{s:preprocessing}, a typical trojan planet will not merely stay fixed at its Lagrangian point, but will oscillate around it in a manner that can be characterized by two superpositioned epicycles, $\Omega_1$ and $\Omega_2$. As derived in \citet{laughlin2002}, these cycles have periods that scale linearly with the orbital period $P_{\rm orb}$, and that otherwise depend only on the masses, in the following ways:

\begin{equation}
P_1 \sim P_{\rm orb} \left( 1 + \frac{27(m_1 + m_2)}{8(m_0 + m_1 + m_2)} \right)
\end{equation}

\begin{equation}
P_2 \sim P_{\rm orb} \sqrt{\frac{4(m_0 + m_1 + m_2)}{27(m_1 + m_2)}}
\end{equation}

where $m_0$ is the mass of the star, $m_1$ the mass of the primary planet, and $m_2$ the mass of the trojan. Since $m_0 \gg m_1 + m_2$, it follows that $P_1$ is close to $P_{\rm orb}$, and that $P_2$ is substantially longer than $P_{\rm orb}$, typically by a factor of $\sim$10--100 for realistic and potentially detectable planetary systems. The amplitude of $\Omega_1$ is significantly smaller than that of $\Omega_2$ (by a factor of $\sim$25, as we will see later).

\begin{figure*}[p]
\centering
\includegraphics[width=16cm]{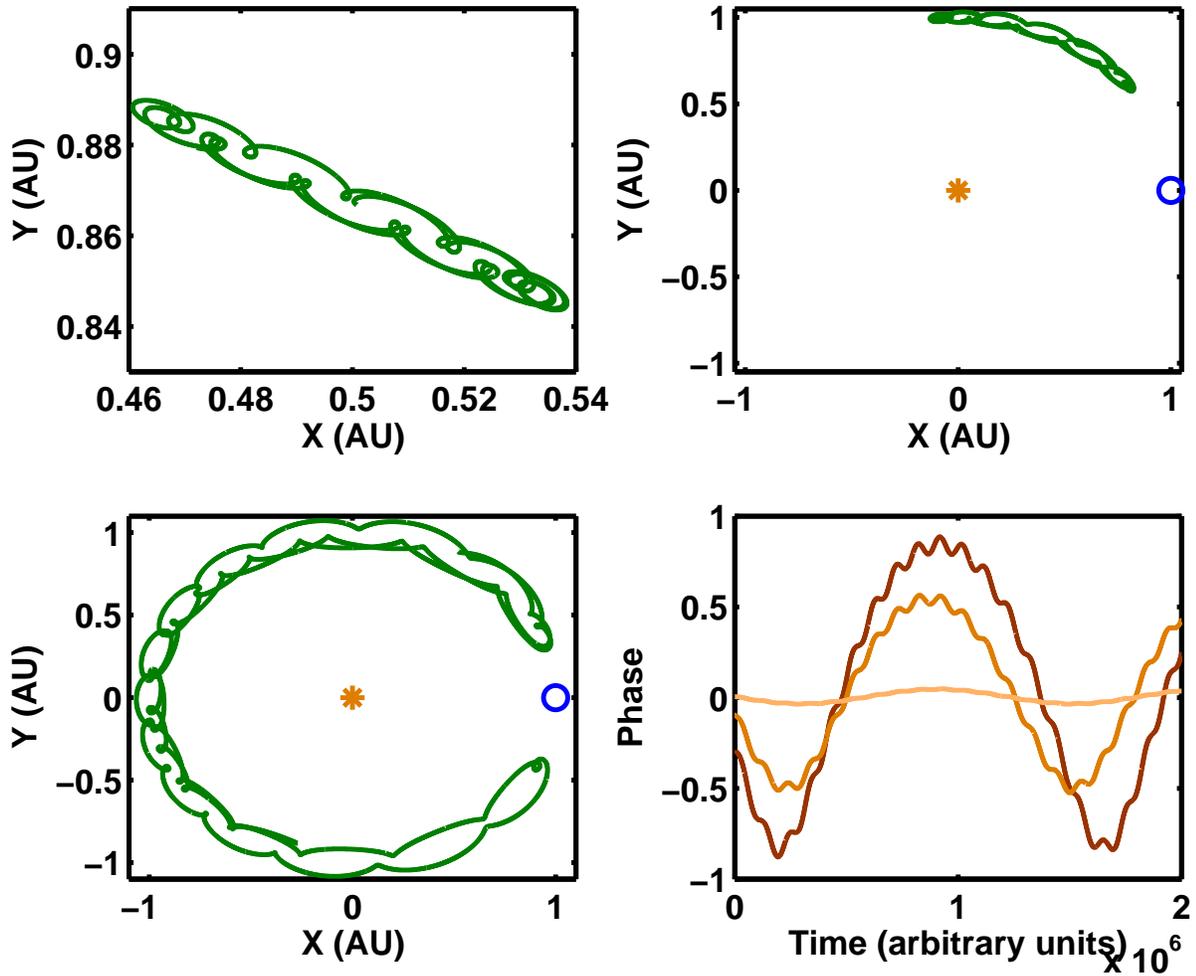}
\caption{A few example trojan orbits, in a frame that co-rotates with the primary planet. Upper left: Zoomed-in view of a fairly small-amplitude trojan orbit. Upper right: Full field view of a large-amplitude tadpole orbit (larger than those considered in the text). The central star is shown as an orange asterisk and the primary planet as a blue ring, for reference. Lower left: A horseshoe orbit. Lower right: Phase variations as a function of time for orbits of three different amplitudes, showing that asymmetries between the peaks and throughs increase with increasing amplitude.}
\label{f:L4orbits}
\end{figure*}

Due to these oscillations, the transit ephemeris of the trojan with respect to the primary planet will vary with time, as the trojan is alternately located ahead of and behind the Lagrangian point around which it librates. In order to be able to systematically detect trojans, we must thus simulate the ephemerides for the full range of epicyclic orbits that the trojan may exhibit. We do this numerically, using the acceleration equation for the full three-body problem adapted from \citet{laughlin2002}:

\begin{equation}
\ddot{\vec{r_2}} = -\frac{G(m_0 + m_2)}{r_2^3} \vec{r_2} - \frac{G m_1}{\Delta ^3} \vec{\Delta} - \frac{G m_1}{r_1^3} \vec{r_1}
\end{equation}

where $\vec{r_1}$ is the vector from $m_0$ to $m_1$, $\vec{r_2}$ is the vector from $m_0$ to $m_2$, and $\vec{\Delta}$ is the vector from $m_1$ to $m_2$. The baseline simulation setup has a time step of $10^{-5}$ years and contains $2 \times 10^{6}$ steps, in a system where $m_0 = 1$~$M_{\rm Sun}$, $m_1 = 1$~$M_{\rm Jup}$, and $m_2 = 1$~$M_{\rm Earth}$, and with a semi-major axis of 1~AU. A few examples of trojan orbit is shown in Fig. \ref{f:L4orbits}, to illustrate the epicyclic motion. We then extract the phase $\phi (t)$ of the trojan in a frame that co-rotates with the primary planet. This is done for a series of simulations with a different initial perturbation $\delta$, where $\delta$ is defined in such a way that if $\delta = 0.001$~AU, then the starting point of the trojan deviates by 0.001~AU from the relevant Lagrangian point in both the x- and y-direction. We vary $\delta$ between 0 and 0.004~AU, where the latter corresponds to a total epicyclic amplitude of 23$^{\rm o}$, roughly equal to the range of Jupiter's trojans. The resulting $\phi (t)$ is fit with a series of sine functions. For small initial perturbations, both $\Omega_1$ and $\Omega_2$ are well represented by pure sine curves, but as the amplitude increases, $\Omega_1$ becomes increasingly asymmetric between its peaks and troughs, as the orbit starts acquiring  the asymmetrically elongated characteristic that has given it the name `tadpole orbit'. We find that this characteristic can be well approximated by adding a co-phased sine wave with $\Omega_2$, with twice the frequency, in addition to a constant offset $A_4$. I.e.:

\begin{equation}
\begin{split}
\label{e:phit}
\phi(t) = A_1 \sin{(\omega_1 t - B_1)} + A_2 \sin{(\omega_2 t - B_2)} \\ 
+ A_3 \sin{(2 \omega_2 t - B_2)} + A_4 
\end{split}
\end{equation}

Here, $B_1$ and $B_2$ are arbitrary phase shifts, and each $A_i$ is a function of $\delta$. We determine these functions by fitting the above relation to the series of simulations with different $\delta$, and fit power laws to the results, since they are well represented by either linear or quadratic fits. We derive the following dependencies:

\begin{equation}
%A_1 = 0.4509 \delta
A_1 = 2.8322 \delta
\end{equation}

\begin{equation}
%A_2 = 8.0168 \delta
A_2 = 50.3711 \delta
\end{equation}

\begin{equation}
%A_3 = -83.5785 \delta^2 - 0.0074 \delta
A_3 = -525.1390 \delta^2 - 0.0467 \delta
\end{equation}

\begin{equation}
%A_4 = 269.2573 \delta^2 - 0.0161 \delta + 0.1667
A_4 = 1691.7934 \delta^2 - 0.1014 \delta + 1.0472
\end{equation}

\begin{figure*}[p]
\centering
\includegraphics[width=16cm]{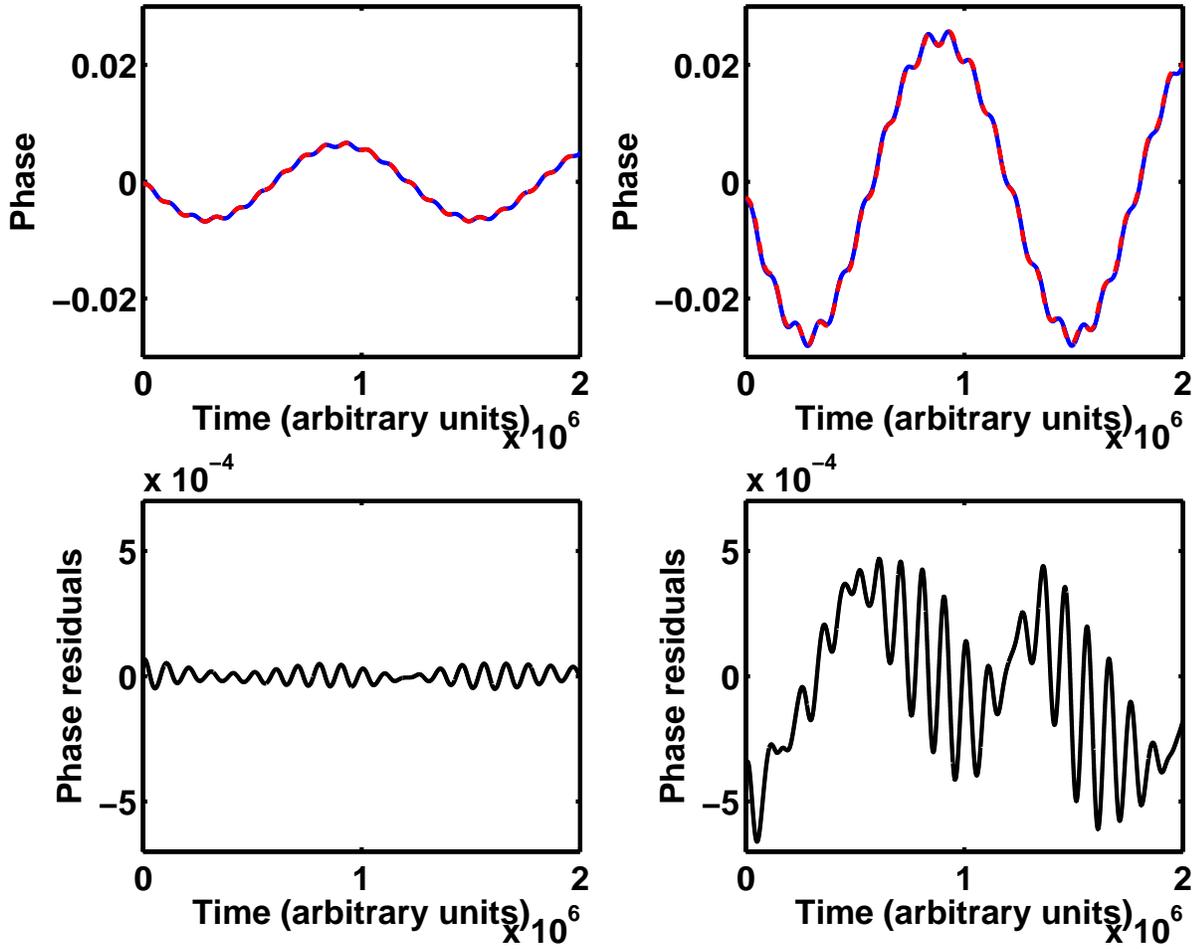}
\caption{Model fits to the phase evolution for two example trojan orbits. Left panels: A relatively small perturbation, $\delta = 0.0008$~AU. Left panels: A relatively large perturbation, $\delta = 0.0032$~AU. Upper panels: Phase evolution with respect to the mean phase as given by the acceleration equation (blue solid lines), compared to our model fits (red dashed lines). The fits are close enough that the curves are difficult to distinguish by eye. Lower panel: Residuals after subtracting the fits. The amplitude of the residuals increases more than the amplitude of the signal between the left and right panels, which is due to the quadratic dependence in the former case versus the linear dependence in the latter.}
\label{f:phase_fit}
\end{figure*}

As can be seen in Fig. \ref{f:phase_fit}, the fits are very good but not perfect, and the standard deviation of the residuals to the fit can be characterized by:

\begin{equation}
%e_{\phi} = 43.4569 \delta^2 - 0.0483 \delta
e_{\phi} = 273.0478 \delta^2 - 0.3034 \delta
\end{equation}

Due to the quadratic dependence, the error is only significant for the largest values of $\delta$. We will discuss the global results of errors on the ephemerides in Sect. \ref{s:results}, but as an illustrative example, if $\delta = 0.004$~AU, the error corresponds to 1 \textit{Kepler} long cadence ($t_{\rm step} = 0.5$~h) if the orbital period is $\sim$42 days ($\sim$2000 $t_{\rm step}$). 

\subsubsection{Searching for Trojans}
\label{s:trojan_search}

The search for signals from potential trojan planets in the data was conducted among all the planet/candidate KOIs using the two complementary approaches of a visual inspection, and a systematic algorithmic search. The visual inspection allows to efficiently discern any false positives and potentially detect unexpected signals or types of orbits that are not actively searched for in our automatic algorithm, such as horseshoe orbits. The algorithmic search, on the other hand, provides the more objective and quantified answer as to whether signals exist in the data or not. All searches were based on the river diagrams explained in Sect. \ref{s:preprocessing}. Our visual inspection did not reveal any convincing trojan candidates, but a few interesting cases of other phenomena are discussed in Sect. \ref{s:individual}.

The idea behind the automatic procedure is the following: For each KOI, we wish to be able to detect any possible corresponding trojan orbit, so long as it is co-inclined with the KOI and has an amplitude of epicyclic motion within that of the trojan swarms of Jupiter (in addition, of course, the trojan has to be of sufficiently large size to be detectable at all in the data). We therefore wish to search a grid of all possible orbits within these constraints, where for a given orbit, a trace is predicted in the river diagram of the KOI, and the would-be transit signal along this trace is summed up. Any orbit among the family of orbits tested that shows a significant feature can then be examined in detail to assess whether or not it is real and to better characterize its properties. 

In practice, the family size of orbits for a given KOI varies in extent, and becomes quite substantial for certain types of objects. There are four parameters that need to be scanned through. The first parameter is the perturbation amplitude, $\delta$. This sets all of the $A_i$ factors in Eqn. \ref{e:phit}. We set the range of this parameter such that it covers the oscillation amplitude range of the Jupiter trojans, and we set the grid steps so that the peak-to-peak amplitude is sampled once for every time it increases by 1~$t_{\rm step}$. This is one of the steps to ensure that the positional error never gets larger than 1~$t_{\rm step}$ for any single transit event. 

The second parameter is a mass parameter, which is essentially determined by $m_1$. The two aggregate masses $m_0+m_1+m_2 \approx m_0$ and $m_1+m_2$ together set the two oscillation periods $P_1$ and $P_2$, as discussed at the top of this section. While the mass of the primary $m_0$ has some considerable uncertainty, it is dwarfed by the uncertainty in $m_1$, since we only know the radius of the (candidate) planet in general, and it can have a wide range of densities. We have furthermore assumed that $m_2 < m_1$, and so $m_1$ provides the most important parameter uncertainty. We therefore fix $m_0$ to the value given in the \textit{Kepler} catalog, set $m_2$ to zero for this purpose, and examine a wide possible range of values for $m_1$. We estimate this range individually for each KOI, based on the estimated planetary radius from the \textit{Kepler}, and the mass-radius relationship of those transiting planets for which both quantities have been measured \citep{weiss2013}. In doing so, we note that while the scatter in the distribution is large, it is not infinite. All objects with a certain radius span a mass range where the lowest and highest mass are within a factor of 10 of each other, apart from in the super-Jovian mass regime where the mass of a $\sim$1~$R_{\rm Jup}$ object can in principle even be in the brown dwarf range, but it is known that such objects are very rare \citep[e.g.][]{grether2006}. Hence, we adopt the mean relationships in \citep{weiss2013}, and establish that for objects with $r_{\rm p} < 1$~$R_{\rm Jup}$, the lower mass limit is $m_{\rm low} = 0.178 r_{\rm p}^{1.89}$, and the upper mass limit is $m_{\rm high} = 1.78 r_{\rm p}^{1.89}$. For objects with $r_{\rm p} > 1$~$R_{\rm Jup}$, we set a lower limit of 0.3~$M_{\rm Jup}$ and an upper limit of 3~$M_{\rm Jup}$. In other words, we try to generously encompass most properties that the planet could have, while simultaneously excluding some parameter space where nothing could realistically reside. With a mass range in hand, we translate it to a distribution in the pair of periods $P_1$ and $P_2$. Within the resulting range, we step through the periods in units of $P_{\rm orb}$, again to ensure that the positional error is smaller than 1~$t_{\rm step}$.

Finally, the third and fourth parameters are simply the phase shifts $B_1$ and $B_2$, which determine which phase each oscillation is in at a given time. We step through $B_2$ in steps of $P_{\rm orb}$ up to $P_2$. $B_1$ is insignificant when $A_1 P_{orb} / 2 \pi$ is smaller than $t_{\rm step}$ which is the case for a rather large range of values of $\delta$, hence its sampling is set to depend on the amplitude in such a way that $4N_{\rm samp}$ uniformly spaced values of $B_1$ are stepped through, where $N_{\rm samp} = A_1 P_{orb} / 2 \pi t_{\rm step}$, rounded downwards to the nearest integer. 

Looping through all of these parameter ranges, for each set of parameters we can generate one specific orbit, which yields a predicted trace in the river diagram, where the central location of each transit is rounded off to the nearest $t_{\rm step}$. Summing up the putative transit signals along this trace, we can make a first assessment of whether a trojan planets exists in this particular orbit or not. In practice, in order to save substantial computational effort, we do this by creating $S/N$ versions of each river diagram prior to running the automatic trojan search. In the $S/N$ diagram, each pixel represents the $S/N$ of an individual hypothetical transit centered on each given $t_{\rm step}$. I.e., for each pixel in the original river diagram, a sum is taken of the signal from $t_{\rm step} - t_{\rm dur}/2$ to $t_{\rm step} + t_{\rm dur}/2$. This sum is divided by the standard deviation of the signal in the local continuum to produce a $S/N$ value at that $t_{\rm step}$. The search algorithm then creates a sum along the trace in the $S/N$ diagram. The resulting value, divided by $\sqrt{N_{\rm per}}$ where $N_{\rm per}$ is the number of rows in the river diagram, gives an estimate of the ensemble $S/N$ for a given trojan orbit. This $S/N$ calculation provides a more robust estimation of the real $S/N$ than a standard root-mean-square estimation, because a strong source of contamination for a given KOI is the presence of transits in the data from other KOIs in multi-planet systems. Such contaminants are (for practical purposes) arbitrarily distributed in phase in a river diagram, and so occasionally one or a few contaminant transits will overlap with the trace of a trojan orbit under investigation. This contamination has a much larger impact on a root-mean-square estimation than our estimation, which favours detection of uniformly distributed events across the trace over strong single-instance events. Nonetheless, our mean-based estimator is still somewhat susceptible to such contaminants, hence we have also used a median-based estimator for the $S/N$, which is calculated as the median of the trace multiplied by $\sqrt{N_{\rm per}}$.

\subsection{Results}
\label{s:results}

\begin{figure*}[p]
\centering
\includegraphics[width=16cm]{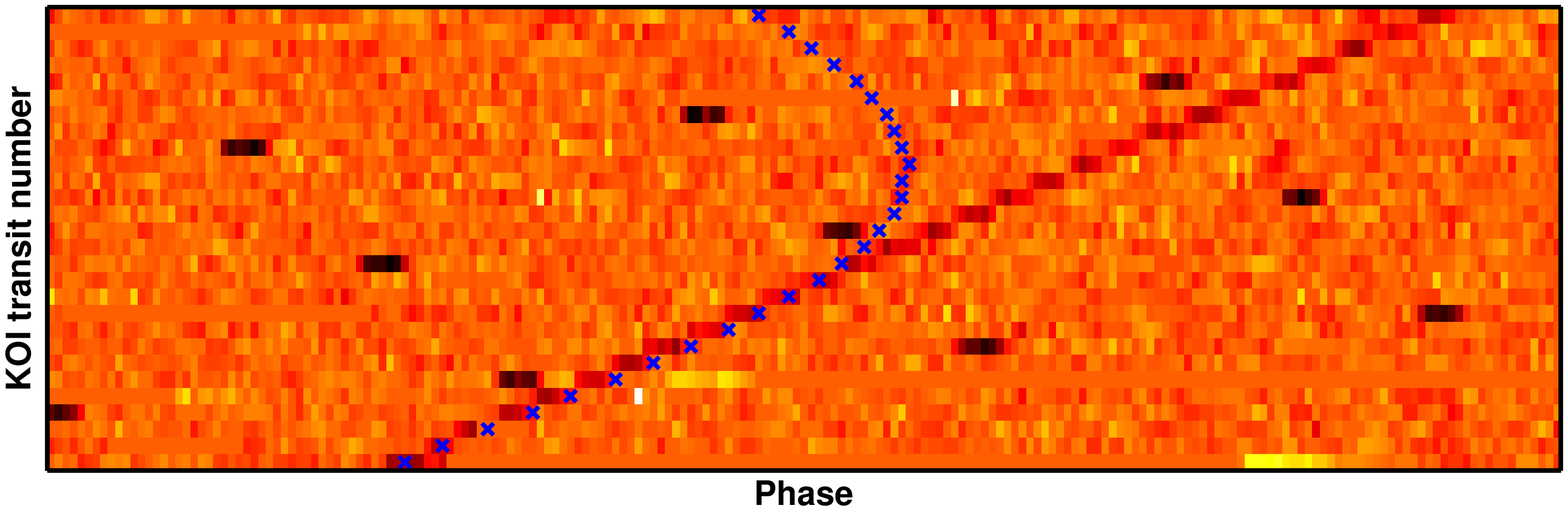}
\caption{Example of a false positive giving a significant signal in the automatic procedure. Shown here is a section of the river diagram of KOI-148.03, in the phase range of $\sim$1.6$\pi$--$1.8 \pi$. In this section we do not see the transit of KOI-148.03 (which is at phase 0.0 by definition), but we see transit features of both KOI-148.01 and KOI-148.02, which both have several times shorter periods than KOI-148.03 and therefore transit several times per row in the river diagram. The deeper transits are from KOI-148.02. The shallower transits are from KOI-148.01, which is close to an integer resonance with KOI-148.03 (8.97:1), and thus forms a relatively mildly diagonal trace across the diagram. The best-fit ephemerides from the automatic procedure are shown as blue crosses. The calculated trojan orbit is able to follow the trace of KOI-148.02 for several consecutive transits, thus summing up to a significant false positive signal.}
\label{f:false_pos}
\end{figure*}

In order to assess whether or not a given target has potential trojan candidates worthy of further study, we set the requirement that the $S/N$ must be $>10$ according to both our mean-based and median-based estimators. This provided 47 leading and 44 trailing candidates, for which we visually examined the path along the river diagram that was determined as statistically significant in our automatic procedure. We found that remaining false positives can explain these cases, and that it is unlikely that any of them correspond to actual trojans. These false positives can be classified in three broad families: (1) KOIs for which there is a significant noise component with a frequency similar to that of the transit signature. The algorithm, which searches for a large number of possible paths along the river diagram, can sometimes find paths with a significant total apparent signal in such cases. (2) KOIs with very long periods, and thus very few rows in the river diagram, such that even the median is not a very robust estimator for the $S/N$. (3) KOIs in multi-KOI systems with near-resonant orbits. In such systems, a river diagram for a planet with a relatively long orbital period will have a number of traces from shorter-orbit periods in it, which run across the entire diagram. If one of these traces happens to fall close to the phases of $\pi / 3$ or $- \pi / 3$, the algorithm will catch them and sometimes be able to fit a significant fraction of the trace. An example of such a case is shown in Fig. \ref{f:false_pos}. 

All of the three above cases can be recognized from the fact that the variance from (putative) transit to transit is much larger than would be expected from the $S/N$ implied by the total signal. The fact that they are limited to a few pathological cases can also be recognized from the fact that many of the leading and trailing false positives occur for the same KOI. Among the 47 leading and 44 trailing candidates, 30 occurred for the exact same KOI, and in many other cases, multiple candidates occurred for different KOIs in the same system.

\begin{figure}[p]
\centering
\includegraphics[width=8cm]{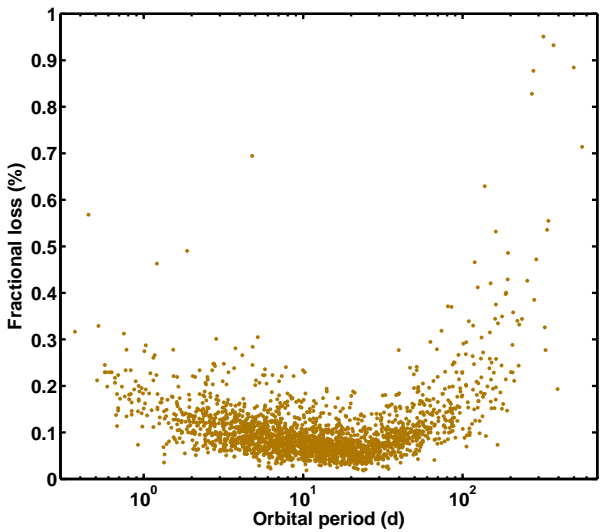}
\caption{Estimated total fractional loss of the transit signal in the systematic trojan search. The typical loss is $\sim$10\%, although for a few cases it is signficiantly larger. The loss fraction increases for larger orbital periods since the impact of the phase error increases, and also increases towards very small orbits since the average transit duration becomes smaller relative to the 0.25~h phase error floor for those cases.}
\label{f:frac_loss}
\end{figure}

\begin{figure}[p]
\centering
\includegraphics[width=8cm]{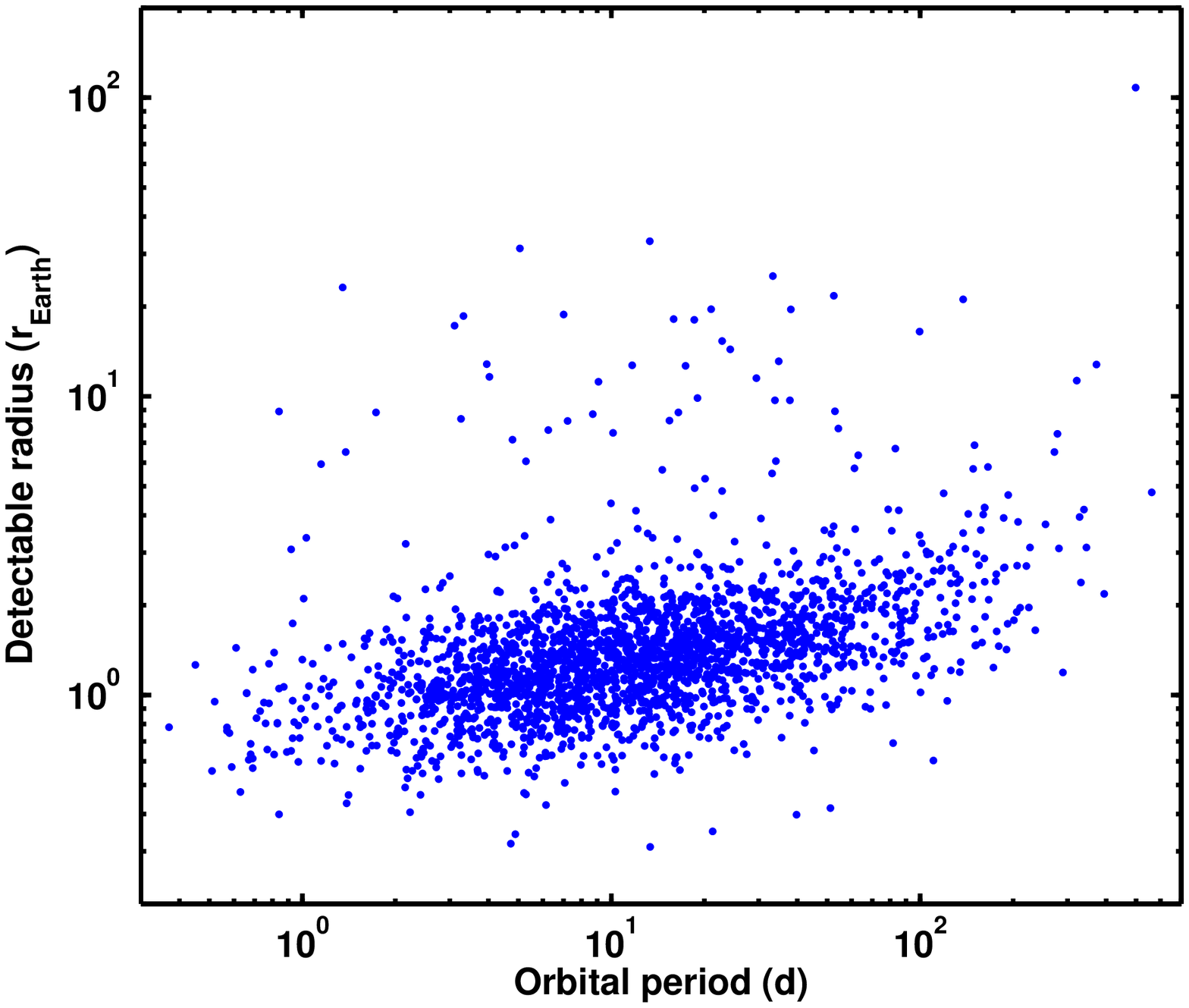}
\caption{Detection limits (10~$\sigma$) for trojan companions to the 2244 KOIs examined in this study. See text for details.}
\label{f:det_radius}
\end{figure}

As a result of this study, we conclude that no trojan companions to the 2244 investigated KOIs were detected at a $>$10~$\sigma$ level. In order to investigate what types of trojans could have been detected if they were present in the data, we calculate what radius a $S/N$ of 10 corresponds to for each potential planet-trojan pair, given the estimated radius of the KOI and its quoted $S/N$ in the \textit{Kepler} catalog. In this process, we take into account the maximum potential losses in our signal from errors in our orbit-fitting procedure. The error $e_{\phi}$ as a function of $\delta$ is given in Sect. \ref{s:orbit}. We calculate it for the maximum applied $\delta$ of 0.004~AU. Furthermore, we set the error such that $e_{\phi} P_{\rm orb} / 2 \pi$ is always greater than 0.25~h, corresponding to the maximal round-off error in the integer-pixel orbit fitting. This error is then related to the duration of the transit, such that if (e.g.) the error is 0.5~h, and the transit duration is 2~h, then 25\% of the total signal is considered lost. The mean calculated loss is 10\%, but is as large as $\sim$90\% in a few individual cases (see Fig. \ref{f:frac_loss}). The temporal error $e_{\phi} P_{\rm orb} / 2 \pi$ of course becomes larger for longer orbital periods, but since the transit duration tends to increase with $P_{\rm orb}$ as well, the loss increases fairly modestly with $P_{\rm orb}$ except for at the very longest periods (Fig. \ref{f:frac_loss}). Adopting these losses, we calculate a set of detectable trojan radii, which are plotted against orbital period in Fig. \ref{f:det_radius}. Typically, the sensitivity is around 1~$R_{\rm Earth}$, in the range of $\sim$0.5--2~$R_{\rm Earth}$, though with significantly worse sensitivities in some cases due to flux losses, or noise, or large orbital periods. We reiterate that these detection limits do not put constraints on the total trojan population, but only to the restricted set of trojan orbits (co-planar, $<$23$^{\rm o}$ libration) that we have been able to systematically probe in this data set.

\subsection{Individual Notes}
\label{s:individual}

In this section, we briefly discuss a few individual cases where a visual inspection of the river diagram turned up unusual features, which according to our literature and archival searches have not yet been reported elsewhere. In addition to these cases, we note that 559.01, 1144.01, 1523.01, 2233.01, and 2470.01 show on-off eclipse patterns -- i.e., they show eclipses in some quarters but not in others. This can be a sign of a blended eclipsing binary false positive signal \citep[e.g.][]{bryson2013}.

\begin{figure*}[p]
\centering
\includegraphics[width=16cm]{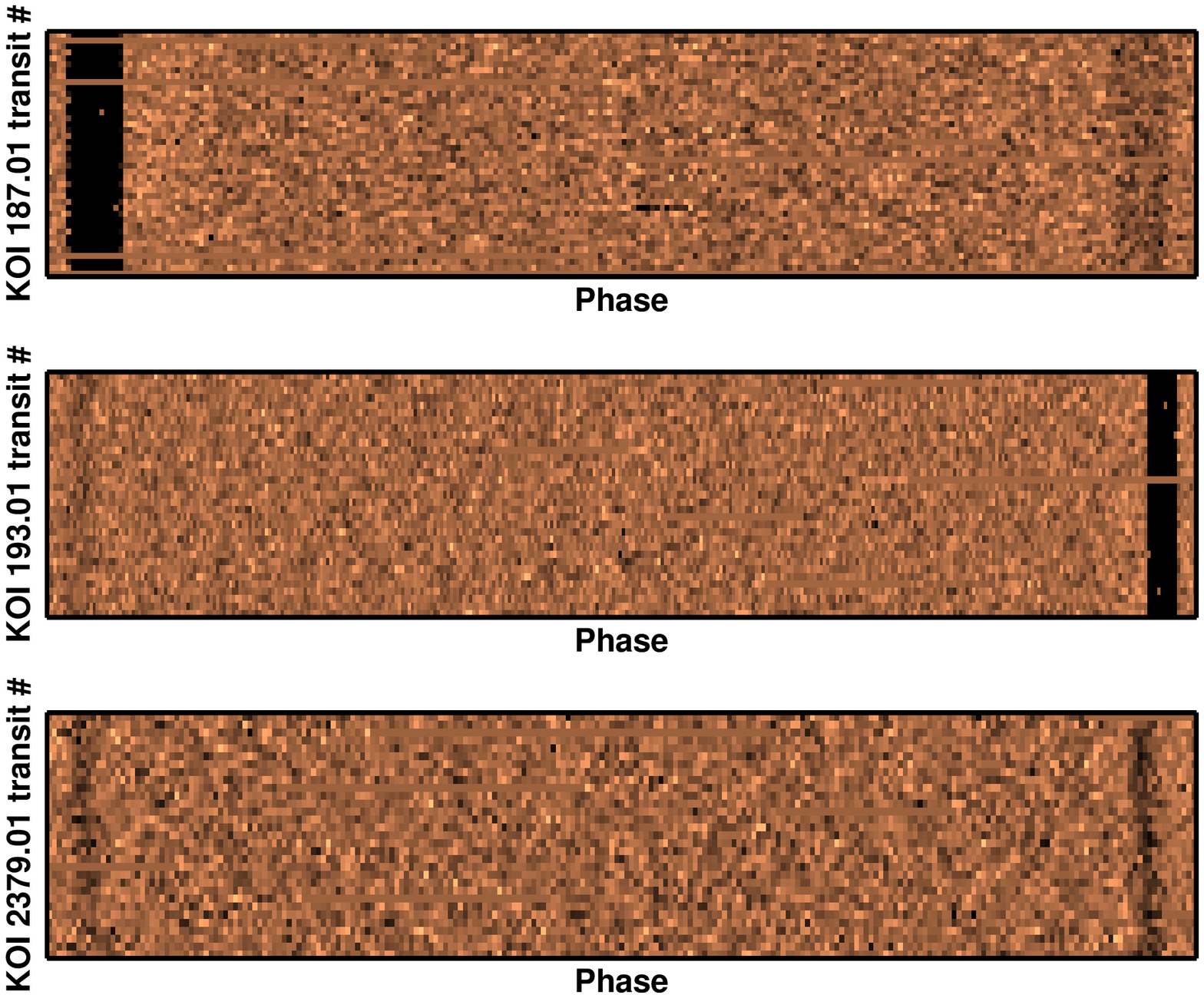}
\caption{River diagrams of KOI-187.01 (upper), KOI-193.01 (middle), and KOI-2379.01 (lower). In each case, the diagram is slightly shifted so that the full primary feature shows and can be compared with the secondary feature. The phase is also cut off just beyond the secondary feature in each case, to provide a reasonable aspect ratio for these long-period cases. For KOI-187.01, the secondary phase is at $\sim$0.3$\pi$, for KOI-193.01 it is at $\sim$1.6$\pi$, and for KOI-2379.01 it is at $\sim$1.8$\pi$, in each case far from the phase of $\pi$ expected for circular orbits.}
\label{f:river_sececl}
\end{figure*}

\textbf{KOI-187.01}: KOI-187.01 has a secondary eclipse feature, strongly offset from a phase of $\pi$. Since the approximate equilibrium temperature of KOI-187.01 (if interpreted as a planet) is 547~K according to \citet{batalha2013}, it is very unlikely that thermal radiation from its surface is related to the secondary eclipse. If the orbit is strongly eccentric (as the offset eclipse indeed indicates), then the temperature in some phases of the orbit will be higher than the equilibrium temperature, but nonetheless, given that the \textit{Kepler} spectral response function cuts off at 900~nm, it would be difficult to get hot enough to contribute substantially to the eclipse depth. Another possibility is reflected radiation if the albedo is high, though this would again require a very eccentric orbit and a fortunate phase of observation. Highly eccentric planets on short orbits are of particular interest for certain hot Jupiter formation scenarios \citep[e.g.][]{socrates2012}, hence this system would be of substantial importance if it were a real planetary system, but false positive scenarios with a blended eclipsing binary seem plausible and should be investigated in detail prior to any conclusions being drawn in this regard. A river diagram of KOI-187.01 (along with diagrams of KOI-193.01 and KOI-2379.01) is shown in Fig. \ref{f:river_sececl}.

\textbf{KOI-193.01}: This is a very similar candidate to KOI-187.01 described above, with a secondary eclipse feature far separated from a phase of $\pi$. Like KOI-187.01, the object would have to be highly eccentric to have any chance of reproducing the secondary eclipse, which would make it highly interesting if it were a planet, but also means that false positive scenarios may be more probable and should be thoroughly investigated.

\textbf{KOI-2379.01}: This is a third case in which a secondary eclipse feature appears far from the $\pi$ phase. The system has a similar period as the other two cases, but for this KOI, the difference between the transit and secondary eclipse features is only a factor $\sim$2. Hence, an eccentric blended binary appears to be a probable cause of this lightcurve.  

\begin{figure*}[p]
\centering
\includegraphics[width=16cm]{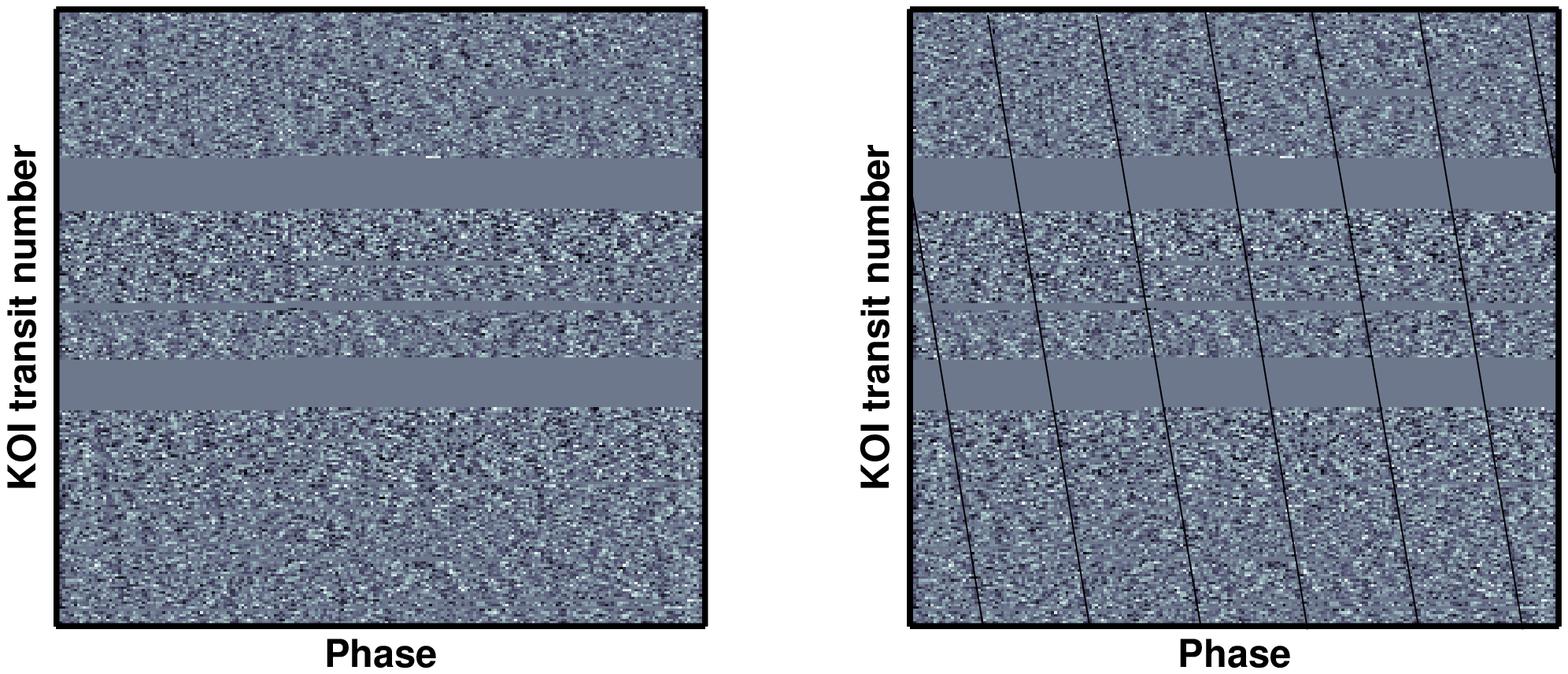}
\caption{River diagram of KOI-2393.01. The transit of the KOI itself is located at phase 0 at the edges of the image. Six faint slightly tilted lines appear across the diagram, indicating the possible presence of an additional planet close to a 1:6 resonance with KOI-2393.01. Left: The river diagram. Right: The same diagram, but with lines drawn across the six traces to guide the eye.}
\label{f:river_6to1}
\end{figure*}

\textbf{KOI-2393.01}: In addition to KOI-2393.01, there are indications of another planet candidate signature in the data, with a period close to 6 times shorter than that of KOI-2393.01 and in the Earth-size range, which to our knowledge has not been reported in the literature. A river diagram is shown in Fig. \ref{f:river_6to1}.

\section{Discussion}

\subsection{KOIs as Possible Trojans}
\label{s:kois_themselves}

In this section, we discuss the possibility that some of the KOIs may not be primary planets (or any common form of false positive), but rather a trojan companion to a larger-non-transiting planet. For this purpose, it is reasonable to exclude all known Kepler multi-planet systems \citep[e.g.][]{fabrycky2013}. As mentioned previously, known trojans in the Solar system have a wide range of mutual inclinations, while the mutual inclinations of Kepler multi-planet systems are small \citep[though non-zero; see e.g.][]{lissauer2011,fang2012}. As a result, it is improbable that a trojan shares the same orbital plane as the other planets in the system, while its primary planet does not. Hence, only systems in which a single KOI has been detected so far are interesting for this discussion. 

If a trojan happens to have a very small $\delta$, then it will undergo negligible TTVs, and thus in principle, almost any detected single KOI could be a trojan companion to an unseen planet. However, there are a few cases that may be particularly interesting in this regard. These are KOIs in single-KOI systems, which undergo short-period periodic TTVs. As one example of this, we will consider the case of KOI~103.01. KOI~103.01 undergoes TTVs with a period of 261.7 days \citep{mazeh2013}, i.e. 17.6 times its orbital period of $P_{\rm orb}$ = 14.9 days . The TTV shape is apparently sinusoidal. This is consistent with a trojan epicyclic motion at the small amplitude of the TTV variation, but not uniquely so, as other planet-planet interactions can cause such behaviour as well. There is no other KOI known in the system, which does however not mean that no other planet is present -- another planet may exist but have a non-equal inclination, or it may have an equal inclination but be located at a larger separation, and fail to transit due to its consequently larger impact parameter. In any case, if we were to interpret the TTVs as epicyclic motion in a 1:1 resonance, then the epicyclic period implies that $m_1 + m_2$ is $\sim$150~$M_{\rm Earth}$. On the other hand, the radius of KOI~103.01 implies that its mass is in the range of 2--14~$M_{\rm Earth}$ \citep[e.g.][]{weiss2013}, i.e. 1--2 orders of magnitude smaller. Hence, if KOI~103.01 were a trojan, it would be dominated in mass by a non-transiting primary planet. 

The benefit of this hypothesis is that it is readily testable. If KOI~103.01 is a planet with a dominant mass in its periodicity range, then its radial velocity (RV) semi-amplitude would be $\sim$0.5--4~m/s, and the RV signal would be in phase with the transit lightcurve. On the other hand, if KOI~103.01 were a trojan, then the RV signal would be dominated by the primary planet, such that the semi-amplitude would be $\sim$40~m/s (assuming a relatively small planet-trojan mutual inclination), and the phase would either lead or trail the transit lightcurve phase by close to 60$^{\rm o}$. The primary star KIC~2444412 is relatively bright at $K_{\rm p} = 12.6$~mag, hence RV follow-up should be readily achievable for this target, at least for testing the trojan hypothesis with its large semi-amplitude prediction. Other similar, but less testable, examples exist, such as KOI~319.01.

Trojan planets with larger libration amplitudes would give unique TTV signatures from their tadpole (or even horseshoe) orbits, but at such amplitudes, they would probably have been missed in the first place by the standard search algorithms. Dedicated quasi-periodic search algorithms, such as QATS \citep{carter2013}, applied to the entire \textit{Kepler} sample would probably have a very good opportunity of finding such objects if they exist at all. If a search is not restricted purely to those systems in which a planet candidate signal has already been detected, as in our survey above, then for each planet-trojan pair that exists, there is an equal probability that either component of the pair transits when the other does not. I.e., if (e.g.) 1\% of all planets have an Earth-sized trojan companion, then there should be of order 10 such trojans already present and detectable in the \textit{Kepler} data, given that the number of detected planet candidates are in the thousands. Hence, a true frequency estimation for trojans in the galaxy could in principle be attained in this manner, although it does seem likely that such a search would be severely computationally demanding.

\subsection{Future Prospects}
\label{s:future}

Like moons, trojans dwell at the bottom of the dynamical hierarchy of planetary systems, but constitute a vital piece of the system architecture, and could potentially play an important role in the context of habitability. As shown by this study and by searches for exomoons also using \textit{Kepler} \citep{kipping2012,kipping2013}, issues related to these types of objects can now start to be quantitatively investigated, and with future facilities, we can expect these opportunities to increase further. For instance, high-contrast imaging techniques are rapidly improving, and with near-future facilities, the planet detection counts are expected to go up significantly \citep[e.g.][]{beuzit2008,macintosh2008,peters2012}. While moons will tend to blend into the point spread function (PSF) of their planet host even for large telescopes, which makes their emitted/reflected radiation hard to distinguish from that of the planet\footnote{Unless they are strongly tidally heated \citep{peters2013}.}, trojans offer the observational advantage of being easily distinguished from the planetary PSF. This means that it would be possible to not only detect (sufficiently large) trojans in images taken for planet-search purposes, but it would also be possible to characterize their atmospheres with follow-up spectroscopy. Of course, the detection limit for a given trojan object in this regard is exactly the same as the detection limit for a planet of the same properties, so direct imaging of, e.g., an Earth-sized trojan in the habitable zone of its star remains out of reach for the near future. In the meantime, however, as has been shown in this study, it will be possible to keep searching for trojans down to Earth-size in \textit{Kepler} data -- potentially even in the habitable zone, at least around late-type stars. 

\section{Conclusions}

In this paper, we have searched for the presence of trojan planets in the \textit{Kepler} data. No convincing cases have been found, despite a sensitivity down to typically 1~$R_{\rm Earth}$ for individual KOIs. However, no broad constraints can be drawn about the trojan population as a whole, in particular because trojans that do not share the same orbital plane as their primary planet are undetectable in the search. We do however note that this also means that, on average, for every system in which a trojan is missed due to it being non-coplanar with the detected KOI, there is an equivalent system in another \textit{Kepler} target system where the trojan plane is transiting from the point of view of Earth, but the planetary plane is not. These objects would be much harder to find in general, since we do not have access to an a proiri period and phase range within which we can limit our search, and also since the number of \textit{Kepler} targets is much larger than the number of KOIs. Typical \textit{Kepler} target trojans would therefore be more computationally demanding to search for, but otherwise leave an equal signal strength and thus (in that sense) have an equal detectability in the data. A systematic search encompassing all \textit{Kepler} targets, rather than the restricted search performed here, would thus be able to provide highly stringent constraints on the frequency of trojans in the Earth-size range and above. On this subject, we have also pointed out that some of the existing individual existing KOIs could be a trojans themselves, dynamically dominated by a primary planet orbiting in a non-transiting plane. A few examples of where this hypothesis is testable are given. 

In summary, we note that any trojan that may exist in the systems that we have examined is either present in the data but masquerading as a regular planet candidate, or is outside of the plane of a genuine planet candidate, or resides in a horseshoe orbit or a tadpole orbit with a very large libration, or has a smaller size than approximately that of Earth.

\acknowledgements
Support for this work was provided by NASA through Hubble Fellowship grant HF-51290.01 awarded by the Space Telescope Science Institute, which is operated by the Association of Universities for Research in Astronomy, Inc., for NASA, under contract NAS 5-26555. This study made use of the CDS services SIMBAD and VizieR, as well as the SAO/NASA ADS service.

\clearpage

\end{document}